\documentclass[nofootinbib,showpacs,pra,preprint,aps,floatfix]{revtex4}
\usepackage{bm}
\usepackage{amssymb}
\usepackage{graphicx}
\usepackage{mathtext}
\usepackage{amsfonts}
\usepackage{mathrsfs}
\usepackage[dvips]{color}
\usepackage{color}

\newcommand{\e}{\mathrm{e}}                  
\newcommand{\ket}[1]{|#1\rangle}             
\newcommand{\bra}[1]{\langle #1|}            
\renewcommand{\d}[1]{\mathrm{d}#1}           
\renewcommand{\v}[1]{\mathbf{#1}}            

\newcommand{\wh}{\widehat}
\begin{document}

\title{Interaction of a two-level atom with a classical field in the context of Bohmian mechanics}

\author{S. V. Mousavi}
\email{s_v_moosavi@mehr.sharif.edu}
\affiliation{Department of Physics, Sharif University of Technology, P. O. Box 11365-9161, Tehran, Iran}
\affiliation{Institute for Studies in Theoretical Physics and Mathematics (IPM), P. O. Box 19395-5531, Tehran, Iran}
\author{M. Golshani}
\email{golshani@sharif.edu}
\affiliation{Department of Physics, Sharif University of Technology, P. O. Box 11365-9161, Tehran, Iran}
\affiliation{Institute for Studies in Theoretical Physics and Mathematics (IPM), P. O. Box 19395-5531, Tehran, Iran}

\begin{abstract}
 We discuss Bohmian paths of the two-level atoms moving in a waveguide through an external resonance-producing field, perpendicular to the waveguide, and localized in a region of finite diameter. The time spent by a particle in a potential region is not well-defined in the standard quantum mechanics, but it is well-defined in the Bohmian mechanics. Bohm's theory is used for calculating the average time spent by a transmitted particle inside the field region and  the arrival-time distributions at the edges of the field region. Using the Runge-Kutta method for the integration of the guidance law, some Bohmian trajectories were also calculated. Numerical results are presented for the special case of a Gaussian wave packet.
\end{abstract}

\maketitle

\section{introduction}
%
%


We study the effect of a classical field on the motion of the ultracold two-level atoms in the context of Bohmian mechanics, taking into account reflection effects, and based on the quantised longitudinal motion. In general, we have to use at least two counter propagating fields to avoid the spatial separation of the internal states due to transverse momentum transfer. However, we follow \cite{SeiMu-EPJD-2007} to direct the atoms in a narrow waveguide and work in a regime for which the  excitation of the transversal modes may be neglected. The field changes the longitudinal atomic momentum as it acts as a quantum mechanical potential. The incident wave packet is divided into two parts once it reaches the field: the reflected one and the transmitted one. We consider the passage of the particles through the field. In the context of Bohmian mechanics \cite{BoI-PR-1952, BoII-PR-1952, HiBo-book-1993, Ho-book-1993, Wy-book-2005, DuGoZa-1992-JSP, DuGoZa-1996-book, Tu-2004-AJP}, the velocity of a particle is given by $\v{v}=\frac{\v{J}}{\rho}$, where $\v{J}=\frac{\hbar}{i 2m} \Bigl(\Psi^{\dagger} \bigtriangledown \Psi - (\bigtriangledown \Psi^{\dagger}) \Psi \Bigr)$ is the current density and $\rho=\Psi^{\dagger} \Psi$ is the probability density. The products on the right hand side of the velocity formula are spinor inner products for the two-component wave function.

 For a one-dimensional scattering problem in which a particle is incident normally (from the left) on the potential $U(\v{r})=V(x) \Theta(x)\Theta(\ell-x)$, three characteristic times are defined: the transmission time $\tau_T(a,b)$ and the reflection time $\tau_R(a,b)$ are the average times spent in the region $a\leq x \leq b$ by transmitted and reflected particles respectively; the dwelling time $\tau_D(a,b)$ is the average time spent between $a$ and $b$ irrespective of whether the incident particle is ultimately transmitted or reflected \cite{Le-SoStCo-1990}. 

 For a brief discussion on the importance of time and time keeping see \cite{MoCaLiPoMu-JP-2008}. There is an ongoing debate on defining the time that characterizes the passage of a quantum particle through a given region. For a recent review see \cite{Wi-PhysRep-2006}. In quantum mechanics, as opposed to classical physics, the meaning of the arrival-time of a particle at a given location is not evident, when the finite extent of the wave function and its spreading becomes relevant, as is the case for cooled atoms dropping out of a trap. Moreover, in the quantum case one expects an arrival-time distribution, and there are different theoretical  proposals for it (see e.g. the review article in \cite{MuLe-PhysRep-2000}, and the book \cite{MuSaEg-book-2002}.) As pointed out by Hannstein et al \cite{HaHeMu-JPB-2005} these arrival-time distributions have been controversial, since they are derived by purely theoretical arguments without specifying a measurement procedure. They provide an operational distribution of arrival times obtained by fluorescence measurements and compare or relate it with more ideal or abstract results.

 Within Bohmian mechanics, the average characteristic times are uniquely defined and each one is obviously a real, non-negative and additive quantity \cite{Le-book-1996}. As Bohm has emphasized in connection with the measurement of the momentum of a particle \cite{BoI-PR-1952, BoII-PR-1952, HiBo-book-1993}, the quantity actually measured may have no relation to its value before the measurement. In Leavens's view \cite{Le-book-1996} the basic difference between the Bohm's trajectory approach and the 'conventional' approach to characteristic times is that Bohmian mechanics consistently treats a 'particle' as an actual particle at all times, not just at those instants when an ideal (i.e. strong) measurement is made of its position.

 As the numerical calculation involved in the calculation of Bohmian trajectories is immense, there have been attempts to compute the Bohmian transmission and reflection times without sampling the trajectories \cite{McLe-PRA-1995, OrMaSu-PRA-1996, Kr-JPA-2005}.

 The next section contains the solution of the stationary Schr\"odinger equation. This is followed by a brief review of the  relevant parts of Bohm's interpretation of quantum mechanics. Sec. \ref{Sec: 4}, contains some numerical calculations. Concluding remarks are made in Sec. \ref{Sec: 5}.


\section{Solution of the stationary Schr\"odinger equation} \label{Sec: 2}

The most common interaction between an atom and an electromagnetic field is of the electric dipole form which couples the  states $\ket{1}$ and $\ket{2}$ of differing parity. The electric dipole approximation is valid when the wavelengths of the driving field are much greater than the mean separation of the electron and the nucleus. As quantum optics is primarily concerned with the interaction between atoms and optical or infrared radiation, this approximation is usually a good one, because the approximation conditions can be met in the laboratory. The atomic dipole operator associated with the two states $\ket{1}$ and $\ket{2}$ has the general form of $\wh{\v{\mu}} = \v{\mu}^* \ket{2}\bra{1} + \v{\mu} \ket{1}\bra{2}$, where the caret~$\wh{~}$~is used to distinguish operators from the corresponding c-numbers. The Hamiltonian describing a two-state atom interacting with a classical electric field $\v{E}(t)$, within the electric dipole approximation, is~\cite{BaRa-book-1997}

\begin{eqnarray} \label{eq: Hamiltonian in Sch picture}
\wh{H} &=& \frac{\wh{p}^2}{2m} + \frac{\hbar}{2} ( \omega_1 \ket{1}\bra{1} + \omega_2 \ket{2}\bra{2}) - \wh{\v{\mu}}.\v{E}~,
\end{eqnarray}
where we have taken $\v{\mu}$ to be real for simplicity. The quantities $\hbar \omega_1$ and $\hbar \omega_2$ are the energies of the states $\ket{1}$ and $\ket{2}$ respectively.
In the interaction picture, and using the rotating-wave approximation, Hamitonian takes the form~\cite{SeiMu-EPJD-2007} 

\begin{eqnarray} \label{eq: Hamiltonian in int picture}
\wh{H} &=& \frac{\wh{p}^2}{2m} - \hbar\Delta \ket{2}\bra{2}+\frac{\hbar}{2}\Omega(\wh{x}) (\ket{2}\bra{1} \e^{-i \phi} + h.c.)~,
\end{eqnarray}
where $\Omega(\wh{x})$ is the position-dependent Rabi frequency, $\phi$ is the phase of the classical field, and  $\Delta=\omega_{L}-\omega_{12}$ denotes the detuning between the field frequency and the atomic transition frequency, $\omega_{12} = \omega_2-\omega_1$, and h.c denotes hermite conjugate. For simplicity we assume that $\Omega(\wh{x}) = \Omega$ for $0 \leq x \leq \ell$, and zero otherwise.
%
%
%
%
%
To obtain the time development of a wave packet incident from the left ($k > 0$), we first solve the stationary equation $\wh{H} \Phi_{k} = E \Phi_{k}$ for the scattering states with the energy $E = \frac{\hbar^2 k^2}{2 m} \equiv E_k$,

\begin{eqnarray} \label{eq:free wave-function}
\Phi_{k}(x) &\equiv& \Phi_{k,R}(x) = \frac{1}{\sqrt{2\pi}}
\left(\begin{array}{c}
\e^{ikx} + R_1(k) \e^{-ikx} \\
R_2(k) \e^{-iqx}
                                            \end{array}\right) ~~~x \leq 0~, \nonumber\\
\Phi_{k}(x) &\equiv& \Phi_{k,T}(x) = \frac{1}{\sqrt{2\pi}}
\left(\begin{array}{c}
T_1(k) \e^{ikx} \\
T_2(k) \e^{iqx}
                                            \end{array}\right) ~~~~~~x \geq \ell~.
\end{eqnarray}
The wavenumber $q$ obeys $q=\sqrt{k^2 + \frac{2m\Delta}{\hbar}}$ due to the conservation of energy, while $R_{1,2}(k)$ and $T_{1,2}(k)$ are reflection and transmission amplitudes for the energy $E_k$. Within the field region, the (unnormalized) dressed state basis that diagonalise the Hamiltonian is given by $\ket{\lambda_{\pm}} = \ket{1} + \frac{2\lambda_{\pm}}{\Omega} \e^{-i\phi} \ket{2}$, where $\lambda_{\pm} = (-\Delta {\pm} \Omega^{\prime})/2$ are the dressed eigenvalues in which $\Omega^{\prime} = \sqrt{\Delta^2 + \Omega^2}$. Thus, the solutions in the interaction region is of the form,

\begin{widetext}
\begin{eqnarray} \label{eq: int wave-function}
\ket{\Phi_k(x)} &=& \frac{1}{\sqrt{2\pi}} \Bigl[\Bigl(A^{(+)}\e^{ik_+ x}+B^{(+)}\e^{-ik_+ x}\Bigr)\ket{\lambda_+}+\Bigl(A^{(-)}\e^{ik_- x}+B^{(-)} \e^{-ik_- x}\Bigr)\ket{\lambda_-}\Bigr]~,
\end{eqnarray}
\end{widetext}
with the wavenumbers $ k_{\pm}^2 = k^2 - 2 m \lambda_{\pm}/\hbar$. The continuity of $\Phi_k(x)$ and its derivative at $x=0$ and $x=\ell$ leads to eight equations with eight unknowns. To impose the matching conditions, it is convenient to use a two-channel transfer matrix formalism~\cite{DaEgHeMu J.Phys.B -2003}.

 For stationary states, the density probability $\rho=\Phi^{\dagger}_k(x) \Phi_k(x)$ is independent of time, and from the continuity equation in 1-dimension, $\frac{\partial \rho}{\partial t} + \frac{\partial J}{\partial x} = 0$, it follows that the current density is a constant. For the left and the right of the field region, the current density is:

\begin{eqnarray}
J_{L} &=& \frac{\hbar k}{m} (1-|R_1(k)|^2)-\frac{\hbar q}{m} |R_2(k)|^2~,\nonumber\\
J_{R} &=& \frac{\hbar k}{m} |T_1(k)|^2 + \frac{\hbar q}{m} |T_2(k)|^2~,
\end{eqnarray}
and from $J_{L}=J_{R}$ one has

\begin{eqnarray}
|R_1(k)|^2 + |T_1(k)|^2 + \frac{q}{k} (|R_2(k)|^2 + |T_2(k)|^2) = 1~.
\end{eqnarray}
The wave packet which describes the particle is of the form

\begin{eqnarray}
\Psi(x,t) &=& \int\d k \Phi_{k}(x) \tilde{\psi}(k) \e^{-i E_k t/\hbar}~,
\end{eqnarray}
where we choose $\tilde{\psi}(k)$ to be 

\begin{eqnarray}
\tilde{\psi}(k) &=& \bigl( \frac{2 \sigma_0^2}{\pi} \bigr)^{1/4} \e^{-\sigma_0^2 (k-k_0)^2}  \e^{-ikx_0}~,
\end{eqnarray}
in which $\hbar k_0$ is the momentum kick as the result of which the particle moves along $x-$axis towards the field, $x_0$ is the centre of the wave packet at $t=0$, and $\sigma_0$ is the initial width of the wave packet.


\section{Sojourn Times in Bohmian mechanics} \label{Sec: 3}

In nonrelativistic Bohmian mechanics the world is described by point-like particles which follow trajectories determined by a law of motion. The evolution of the positions of these particles is guided by a wave function which itself evolves according to the Schr\"odinger's equation. Bohmian mechanics makes the same predictions as the ordinary nonrelativistic quantum mechanics for results of any experiment, provided that we assume a random distribution for the initial configuration of the system and the apparatus in the from of $\rho(x,0)= \Psi^{\dagger}(x,0) \Psi(x,0)$. If the probability density for the  configuration satisfies $\rho(x,t_0)= \Psi^{\dagger}(x,t_0) \Psi(x,t_0)$ at some time $t_0$, then the density to which this is carried by the continuity equation at any time $t$ is also given by $\rho(x,t)= \Psi^{\dagger}(x,t) \Psi(x,t)$~\cite{Bo-PR-1953}. 
Because most quantum measurements boil down to position measurements, Bohm's theory and the standard quantum mechanics will in general yield the same detection probabilities. The situation is different, however, if one considers, for example, measurements involving time related quantities, such as time of arrival, tunnelling times etc. Bohm's theory makes unambiguous predictions for such measurements, but in the conventional quantum mechanics there is no consensus about what these quantities should be ~\cite{St-quant-2005}.
Given the initial position $x^{(0)} \equiv x(t=0)$ of a particle with the initial wave function $\v{\Psi}(x, t=0)$, its subsequent trajectory $x(x^{(0)}, t)$ is uniquely determined by the simultaneous integration of the time dependent Schr\"odinger equation, TDSE, and the guidance equation $\frac{dx(t)}{dt}=v(x(t),t)$. 

 As pointed out by Leavens, the noncrossing property of Bohmian trajectories $x(x^{(0)}, t)$ leads to a very important consequence in 1-dimension: there is a critical trajectory (starting at $x_c^{(0)}$) that divides the wave packet into two parts, the transmitted one ($x^{(0)} > x_c^{(0)}$) and the reflected one ($x^{(0)} < x_c^{(0)}$) ~\cite{Le-PhyslettA-1993}. The reflection-transmission bifurcation curve $x_c(t)$ separating the transmitted trajectories from the reflected ones is given by 

\begin{eqnarray} \label{eq: trans-prob}
|T|^2 &=& \displaystyle\int_{x_c(t)}^\infty \d x \Psi^{\dagger}(x,t) \Psi(x,t)~,
\end{eqnarray}
where $|T|^2$ is the transmission probability which is defined as $\displaystyle\lim_{t \to \infty} \displaystyle\int_{\ell}^\infty \d x \Psi^{\dagger}(x,t) \Psi(x,t)$ and could be calculated from the stationary-state transmission probabilities $|T_1(k)|^2$ and $|T_2(k)|^2$ as

\begin{eqnarray} \label{eq: trans-prob versus stationary ones}
|T|^2 &=& \sqrt {\frac{2 \sigma_0^2}{\pi}} \displaystyle\int_{-\infty}^\infty \d k  \nonumber\\
&\times& \Bigl( |T_1(k)|^2 + \frac{q}{k} |T_2(k)|^2 \Bigr) \e^{-2\sigma_0^2 (k-k_0)^2}~,
\end{eqnarray}

 The dwelling time $\tau_D(0,\ell)$ in the field region is given by

\begin{eqnarray} \label{eq: dwelling-time}
\tau_D(0,\ell) &=& \displaystyle\int_{0}^\infty \d t \displaystyle\int_{0}^\ell \d x \Psi^{\dagger}(x,t) \Psi(x,t)~.
\end{eqnarray}
Inserting $1 \equiv \Theta(x-x_c(t)) + \Theta(x_c(t)-x)$ into the integrand of this equation, one gets

\begin{eqnarray} \label{eq: weighting relation}
\tau_D(0,\ell) &=& |T|^2 \tau_T(0,\ell) + |R|^2 \tau_R(0,\ell)~,
\end{eqnarray}
in which

\begin{eqnarray}
\tau_T(0,\ell) &=& \frac{1}{|T|^2} \displaystyle\int_{0}^\infty \d t \displaystyle\int_{0}^\ell \d x \Psi^{\dagger}(x,t) \Psi(x,t) \Theta(x-x_c(t))~,\nonumber\\
\end{eqnarray}

\begin{eqnarray}
\tau_R(0,\ell) &=& \frac{1}{|R|^2} \displaystyle\int_{0}^\infty \d t \displaystyle\int_{0}^\ell \d x \Psi^{\dagger}(x,t) \Psi(x,t) \Theta(x_c(t)-x)~,\nonumber\\
\end{eqnarray}
where $\tau_D(0,\ell)$, $\tau_T(0,\ell)$ and $\tau_R(0,\ell)$ are, respectively, the mean dwelling, transmission and reflection times respectively, and $|T|^2$ and $|R|^2 = 1 - |T|^2$ are transmission and reflection probabilities, respectively. Within the Bohmian mechanics, the average characteristic times $\tau_{X=D,T,R}(0,\ell)$ are uniquely defined and each one is obviously a real, non-negative and additive quantity in the sense of measures: $\tau_X(x_1,x_3) = \tau_X(x_1,x_2) + \tau_X(x_2,x_3)$, with $x_1 \leq x_2 \leq x_3$, in which $X$ could be $D$, $T$ or $R$~\cite{Le-book-1996}. 
%

The total presence probability at the right of the point $x$ at time $t$, $Q(x, t)$, is given by

\begin{eqnarray}
Q(x, t) &=& \displaystyle\int_{x}^\infty \d x^{\prime} \Psi^{\dagger}(x^{\prime},t) \Psi(x^{\prime},t) = \displaystyle\int_{0}^t\d t^{\prime} J(x, t^{\prime})~.\nonumber\\
\end{eqnarray}
 As discussed by Oriols, Martin and Su\~n\'e~\cite{OrMaSu-PRA-1996}, the problem of the average characteristic times can be reduced to the calculation of the total presence probability at the right of the edges of the field region, $Q(0, t)$ and $Q(\ell, t)$:

\begin{eqnarray} \label{eq: tau_D}
\tau_D &=& \displaystyle\int_{0}^\infty \d t \Bigl(Q(0,t) - Q(\ell,t) \Bigr)~,
\end{eqnarray}

\begin{eqnarray} \label{eq: tau_T}
\tau_T &=& \frac{1}{|T|^2} \displaystyle\int_{0}^\infty \d t \Bigl[ min \Bigl( Q(0,t), |T|^2 \Bigr) - Q(\ell, t) \Bigr]~,
\end{eqnarray}
 and

\begin{eqnarray} \label{eq: tau_R}
\tau_R &=& \frac{1}{|R|^2} \displaystyle\int_{0}^\infty \d t \Bigl[ max \Bigl( Q(0,t), |T|^2 \Bigr) - |T|^2 \Bigr]~.
\end{eqnarray}


 In the causal interpretation, the distribution of arrival-times $P_{b}(t)$ at a given interface $x=b$, with $b \geq \ell$, is related to the current density $J(b,t)$ by 

\begin{eqnarray} \label{eq: arrive-pos}
P_{b}(t) &=& \frac{J(b,t)}{|T|^2} = \frac{1}{|T|^2} \frac{\partial Q(b, t)}{\partial t}~,
\end{eqnarray}
provided there are no re-entrant trajectories through $x=b$~\cite{Le-PhyslettA-1993, McLe-PRA-1995}. This expression is valid in the absence of an arrival-time detector, because an arrival-time detector is not included in the Hamiltonian to derive this expression. Thus, this is an expression for the ideal or intrinsic arrival-time distribution~\cite{MuSaEg-Le-book-2002}. If $Q(b, t)$ is not monotonic, then $J(b, t)$ must be negative for some times, implying re-entrant trajectories through $b$. In this case the arrival-time distribution is given by

\begin{eqnarray} \label{eq: arrive-neg}
P_{b}(t) &=& \frac{|J(b, t)|}{\displaystyle\int_{0}^\infty \d t |J(b, t)|}~,
\end{eqnarray}
allowing for multiple arrival-times, both from the left and right, associated with any re-entrant trajectory that might occur~\cite{McLe-PRA-1995, OrMaSu-PRA-1996, Le-book-1996, MuSaEg-Le-book-2002}.

%


\section{Numerical Results} \label{Sec: 4}

 Consider Cs atoms with mass $m=2.2 \times 10^{-25}$kg, impinging on a classical on-resonance $\frac{\pi}{2}-$pulse, with the Rabi-frequency $\Omega=(n+1/2) \times \frac{v_0}{\ell}$, where $n$ is an integer number. The initial velocity of the particle is chosen to be $v_0=0.01$m/s, the width and the phase of the field is chosen as $\ell=10^{-4}$m and $\phi=0$  respectively, and $n=650$ and $\sigma_0=20\mu$m. With these values, the fraction of the  interaction energy to the kinetic energy of the particle is $\frac{\hbar \Omega}{E_k} = 1.96$. 

\begin{figure}
\includegraphics[width=10cm,angle=0]{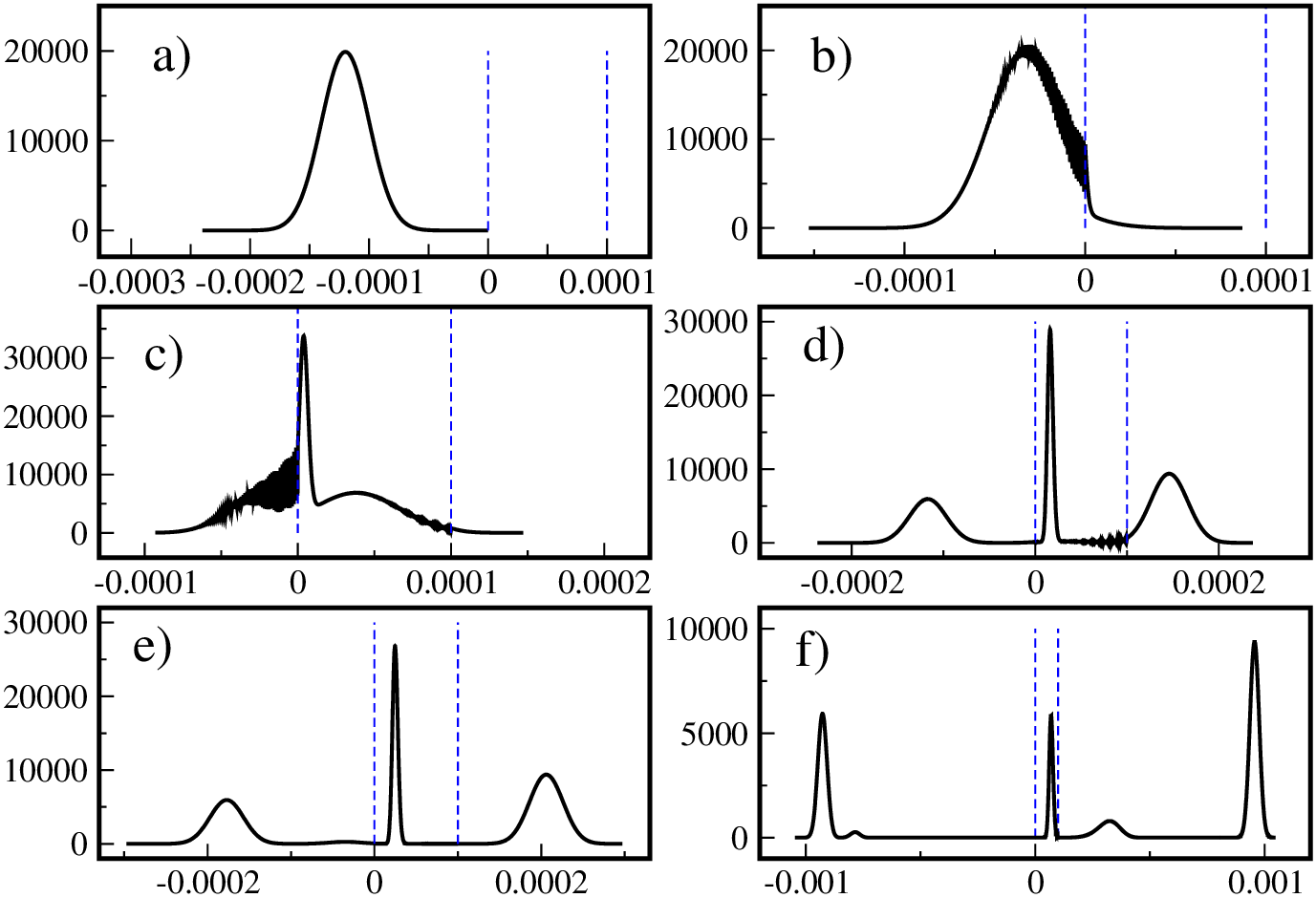}
\centering
\caption{Probability density versus distance $x$(m) at times a)~$t=0$ ms, b)~$t=8.7$ ms, c)~$t=14.7$ ms, d)~$t=23.7$ ms, e)~t=$29.7$ ms and f)~$t=104.7$ ms. Vertical dashed lines show edges of the field.}
\vspace{0.7cm}
\label{fig: density}
\end{figure}

 Fig.~\ref{fig: density} shows the probability density for a wave packet approaching the field region and colliding with the field, and also the construction of the transmitted and reflected wave packets. This figure resembles the behaviour of a potential well or a potential barrier (with a height smaller than the kinetic energy of the arriving particle), when a wave packet approaches the potential region.



\begin{figure}
\centering
\includegraphics[width=10cm,angle=0]{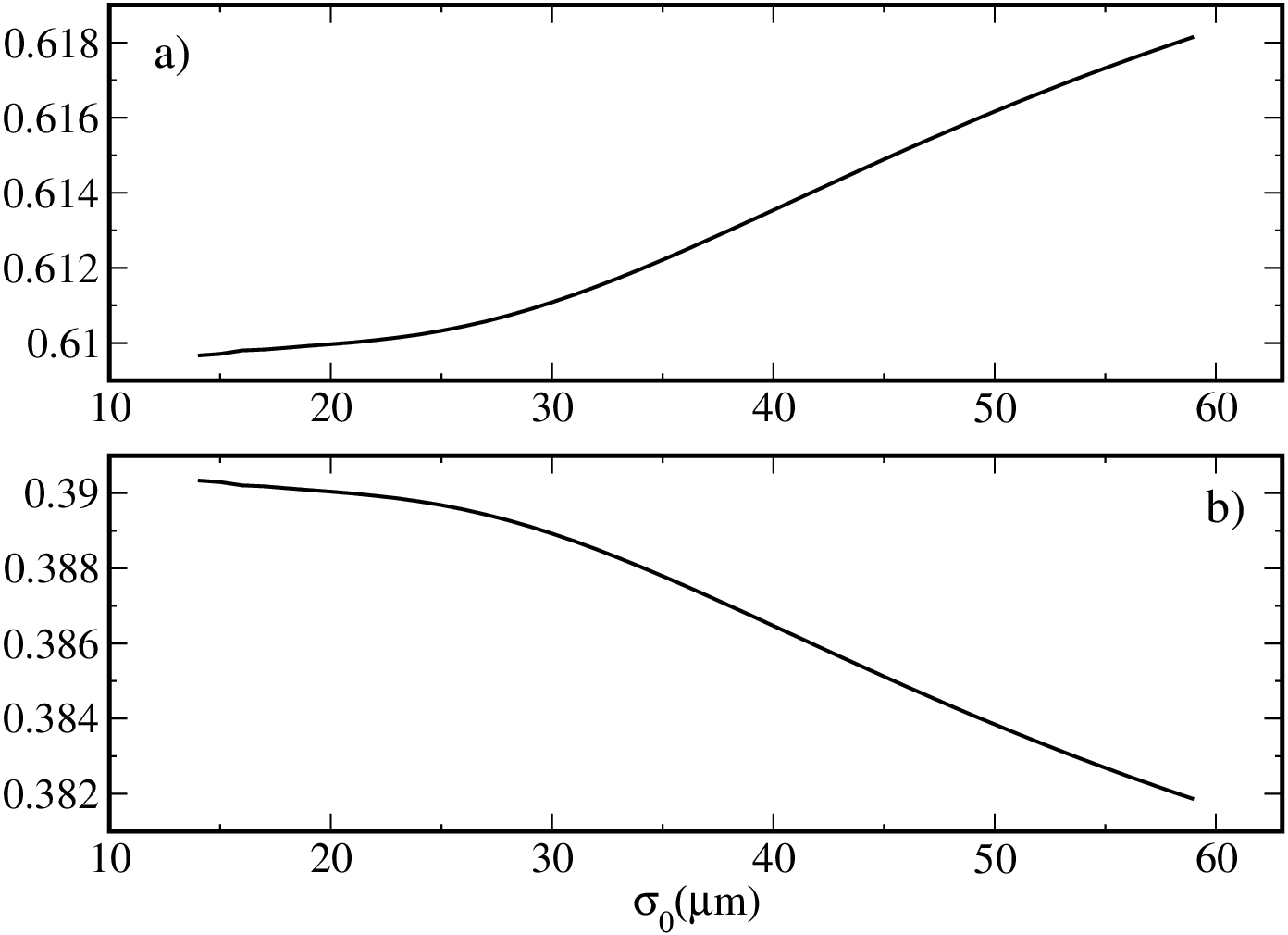}
\caption{a) Transmission and b) reflection probability versus $\sigma_0 (\mu m$).}
\label{fig: trans-sigma}
\end{figure}

\begin{figure}
\centering
\includegraphics[width=10cm,angle=0]{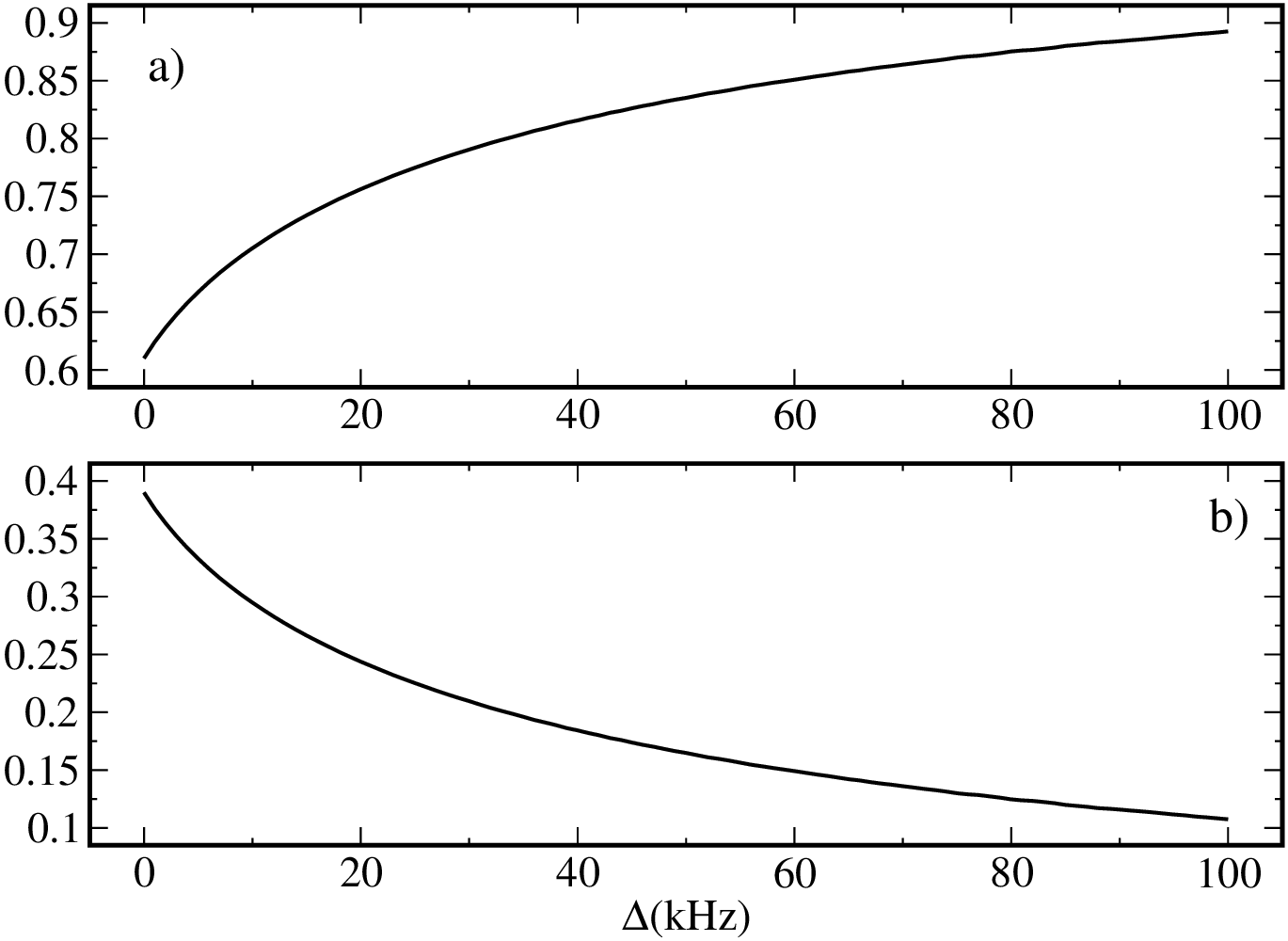}
\caption{a) Transmission and b) reflection probability versus $\Delta$ (kHz).}
\label{fig: trans-Delta}
\end{figure}

 Numerical calculations show that transmission probability $|T|^2$ increases with both $\sigma_0$ and $\Delta$ (Fig.~\ref{fig: trans-sigma} and Fig.~\ref{fig: trans-Delta}). Increasing $\Delta$ means that detuning is increased, thus the interaction between the particle and the field reduces, i.e., transmission probability is increased.
In the case that total energy of the atom, the kinetic energy of the excited state and the interaction energy $\hbar \Omega$ are negligible with respect to $\hbar \Delta$, one can show, using the stationary Schr\"odinger equation, that the atom propagates freely in the ground state.

\begin{figure} 
\centering
\includegraphics[width=10cm,angle=0]{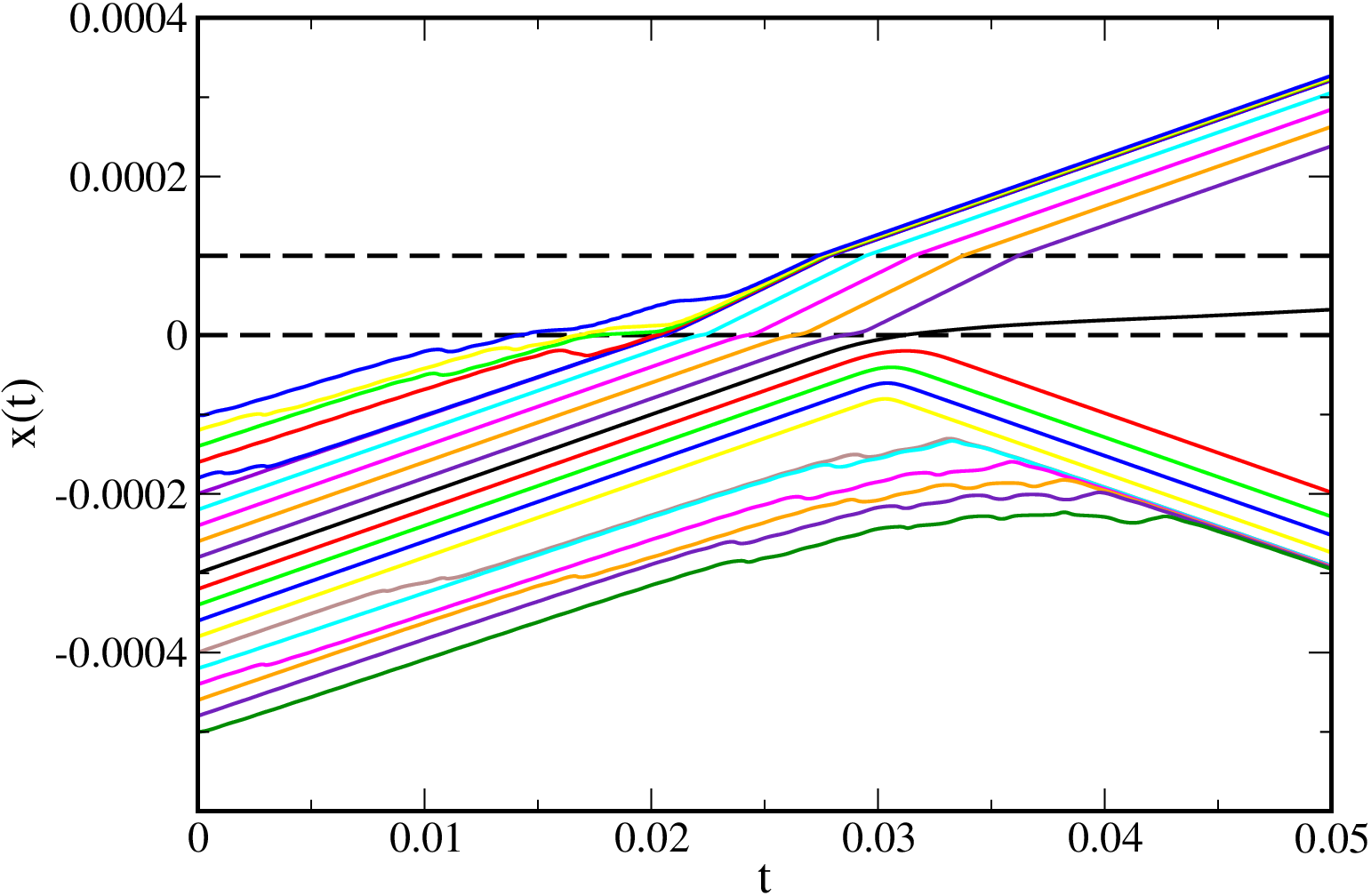}
\caption{A selection of Bohmian paths. The position of the field is indicated by the horizontal dashed lines. The black curve, which starts at $x_c^{(0)} = -0.126$ mm, is the bifurcation trajectory.}
\label{fig: trajectory}
\end{figure}

 In Fig.~\ref{fig: trajectory}, a selection of Bohmian trajectories in the neighbourhood of the transmission-reflection bifurcation is shown, using the Runge-Kutta method for the integration of the guidance equation. From Eq.~(\ref{eq: trans-prob versus stationary ones}), we get $|T|^2 = 0.61$ for $\sigma_0=20\mu$m. Since the initial wave packet is an even function of $x$ around $x_0$, the centre of the initial wave packet, the total probability at the right and left of $x_0$ is each one equal to $0.5$. Thus, Eq.~(\ref{eq: trans-prob}) at $t=0$ and $|T|^2 = 0.61$ yields  $x_c^{(0)}=x_0 - 0.56 \times 10^{-05} < x_0$, i.e., the bifurcation trajectory starts in the back half of the initial wave packet. At points where the effective force (the classical and the quantum ones) is small, or equivalently probability density  is large, trajectories congregate and give rise to fringes, a matter-wave analogue of Wigner's optical fringes ~\cite{Ho-book-1993}. It should be noted that the continuity of the wave-function and its derivative at the boundaries, imply that probability density and the probability current density both are continuous functions at the boundaries. Thus, it is found from the guidance equation that the velocity of a Bohmian particle is continuous at the boundaries. This point is not evident from Fig.~\ref{fig: trajectory} because of its scale, but it was found by zooming on this figure at the boundaries.


 Using Eqs.~(\ref{eq: tau_D}, \ref{eq: tau_T} and \ref{eq: tau_R}), one gets $\tau_D=40$ ms, $\tau_T=39$ ms and $\tau_R=42$ ms for $\sigma_0=20$ micron. These values satisfy the weighting relation $\tau_D = |T|^2 \tau_T + |R|^2 \tau_R$, which is  an unprovable relation in the standard quantum mechanic. Eq.~(\ref{eq: weighting relation}) shows that the integrated density under the field region consists of additive contributions from the transmitted and from the reflected beams. As pointed out by Landauer and Martin ~\cite{LaMa-Rev.mod.Phys-1994}, one does not sum probabilities in quantum mechanics; rather, it is the complex amplitudes that are summed over. Eq.~(\ref{eq: weighting relation}) neglects the possibility of interference between the amplitudes for reflection and transmission.
 
\begin{figure}
\centering
\includegraphics[width=10cm,angle=0]{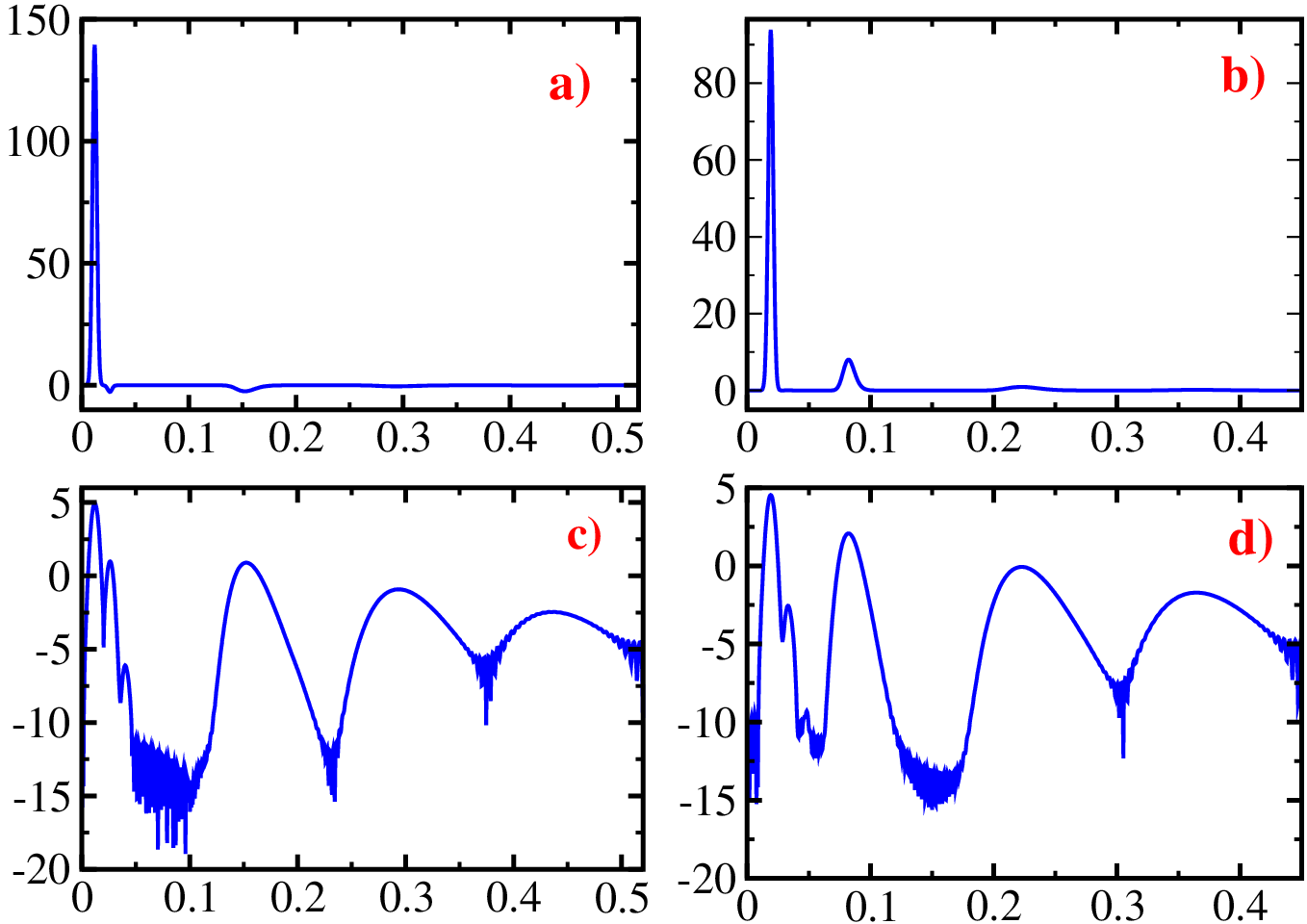}
\caption{ a)~$J(0, t)$, b)~$J(\ell, t)$, c)~$\log(|J(0, t)|)$ and d)~$\log(J(\ell, t))$ versus time(s). From a) and c) one can find that in a) there is a big positive peak and then five small negative ones; the first represents the wave packet as it enters the field region, the later five represent the reflected packets. From b) and d) it's concluded that there are five positive peaks in b).}
\label{fig: current}
\end{figure}

\begin{figure}
\centering
\includegraphics[width=10cm,angle=0]{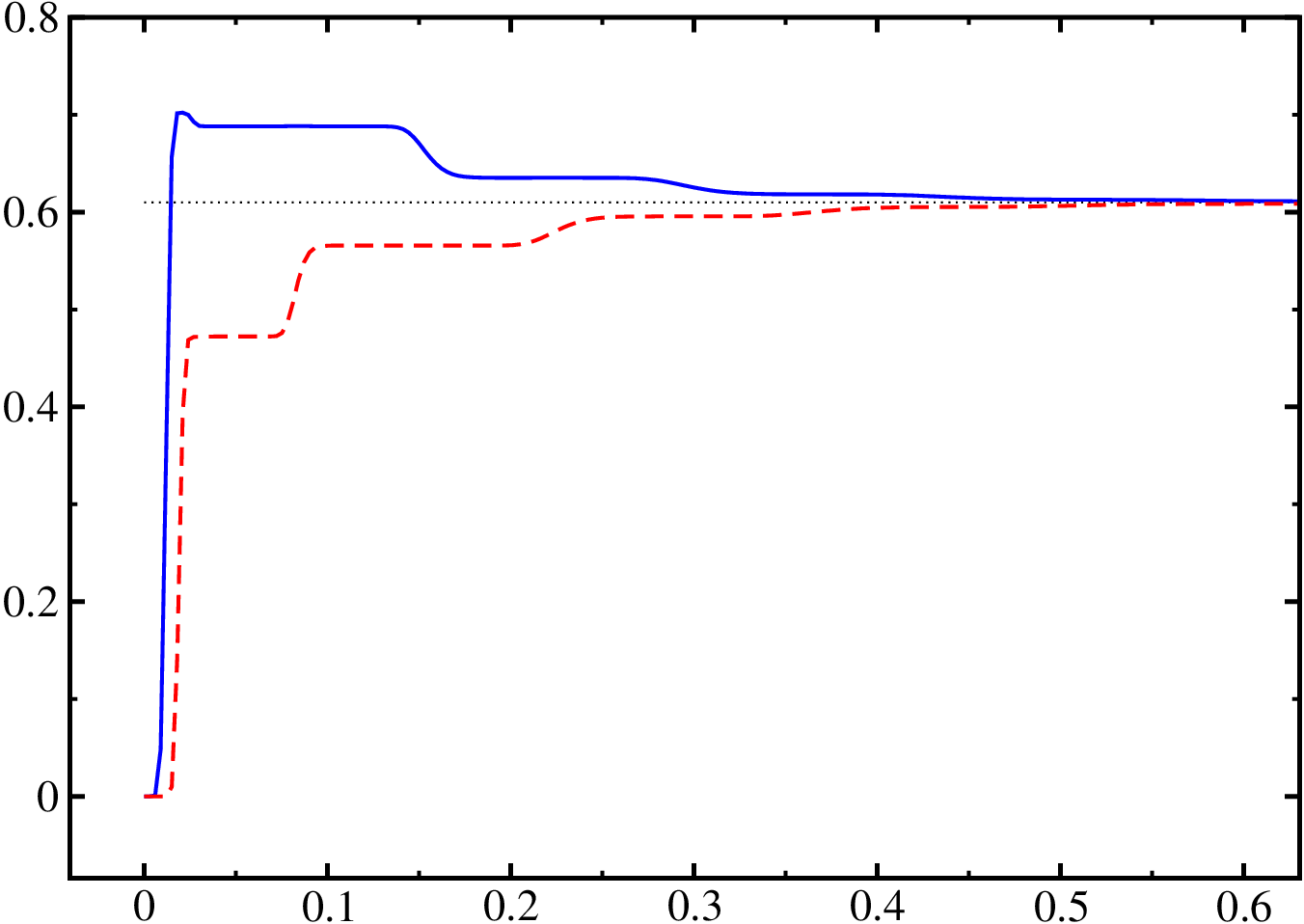}
\caption{ $Q_0(t)$ (solid line) and $Q_{\ell}(t)$ (dashed line) versus time(s). The horizontal dotted line shows the  transmission probability $|T|^2$.}
\label{fig: presence probability}
\end{figure}

\begin{figure} 
\centering
\includegraphics[width=10cm,angle=0]{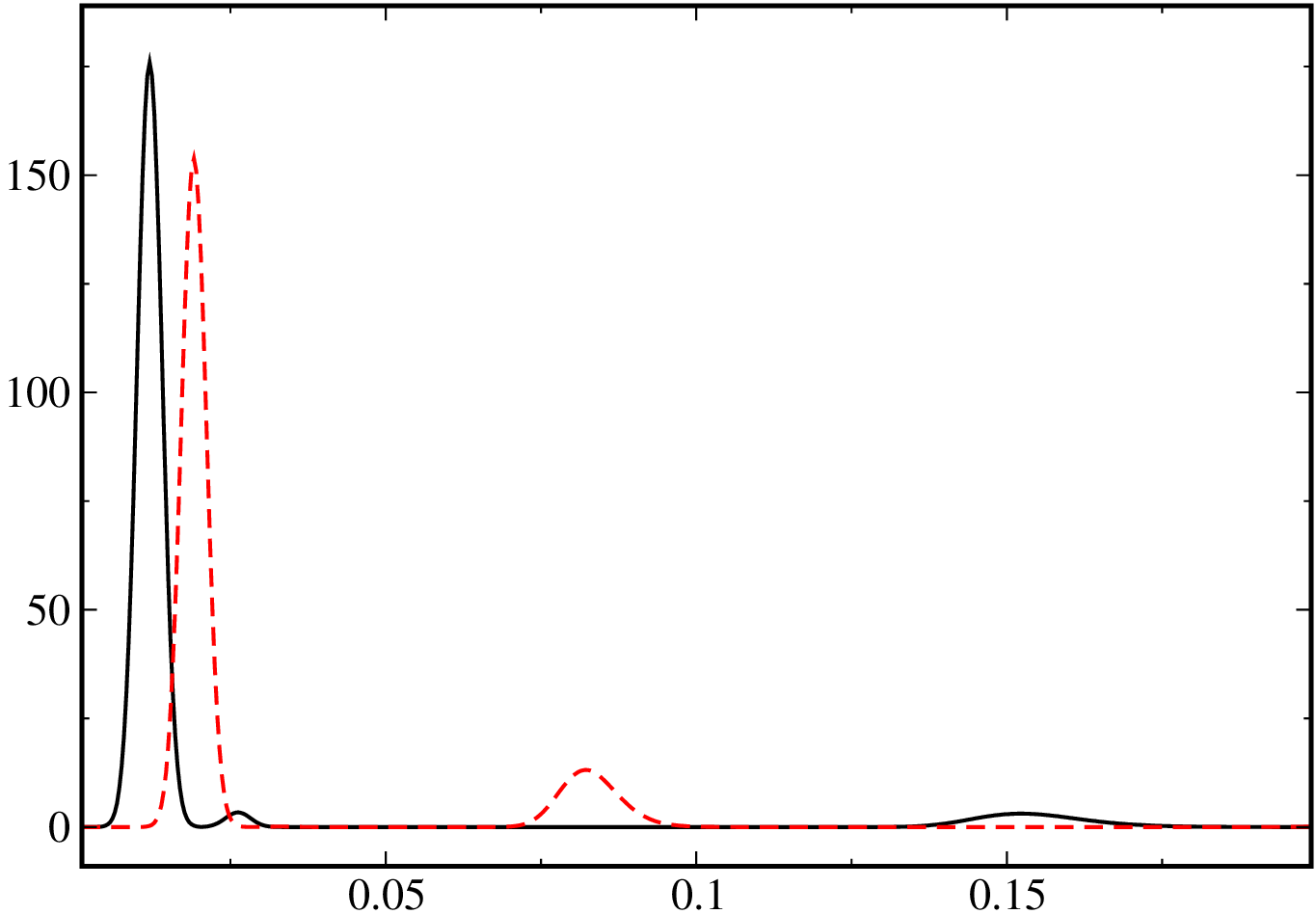}
\caption{ $P_0(t)$ (solid line) and $P_{\ell}(t)$ (dashed line) versus time(s).} 
\label{fig: arriv}
\end{figure}

 Fig.~\ref{fig: current} shows the current and the logarithm of the current densities and Fig.~\ref{fig: presence probability} shows the  corresponding integrated current densities, $Q(0, t)$ and $Q(\ell, t)$, at the edges $x=0$ and $x=\ell$; as a function of time. We have plotted the logarithm of the current densities to see the number of peaks of the  current densities. Fig.~\ref{fig: current} shows that $J (0, t)$ is negative at some instants, corresponding to some re-entrant trajectories at this boundary, but $J ({\ell}, t)$ is positive for all the times. 
Finally, Fig.~\ref{fig: arriv} shows the arrival-time distributions at $x=0$ and $x=\ell$, according to Eq.~(\ref{eq: arrive-pos}) and Eq.~(\ref{eq: arrive-neg}) respectively. 

%


\section{Discussion} \label{Sec: 5}
 The calculations presented here in the context of Bohmian mechanics, are for the traversal time, the dwelling time and the arrival-time distributions in the absence of any measuring device. Hence, before comparison with experiment can be seriously considered, a method for doing the measurement must be specified and the effect of the interaction between the propagating particle and the apparatus must be included in the calculation. As Bohm emphasized, the quantity actually measured may have no relation to the quantity before the measurement. An important feature of the causal interpretation is that, since it gives a well-defined, unambiguous and unique prescription for calculating the characteristic times and the arrival-times, it is refutable experimentally, provided that the Bohmian trajectory calculation of the desired quantity and the corresponding experiment can both be carried out with sufficient accuracy. On the other hand, since there is no prescription in the basic tenets of the conventional interpretation for calculating the characteristic times, that interpretation isn't refutable via comparison of the calculated and the actual experimental characteristic times~\cite{Le-SoStCo-1990}.

\section{Acknowledgment}

We are very indebted to R. Tumulka for helpful discussions and corrections and also S. Goldstein and H. Nikolic for valuable suggestions. S. V. M. is grateful to J. G. Muga for providnig useful information.

 \pagebreak

\textbf{Figure captions}

Figure 1: Probability density versus distance $x$(m) at times a)~$t=0$ ms, b)~$t=8.7$ ms, c)~$t=14.7$ ms, d)~$t=23.7$ ms, e)~t=$29.7$ ms and f)~$t=104.7$ ms. Vertical dashed lines show edges of the field.

Figure 2: a) Transmission and b) reflection probability versus $\sigma_0 (\mu m$).

Figure 3: a) Transmission and b) reflection probability versus $\Delta$ (kHz).

Figure 4: A selection of Bohmian paths. The position of the field is indicated by the horizontal dashed lines. The black curve, which starts at $x_c^{(0)} = -0.126$ mm, is the bifurcation trajectory.

Figure 5: a)~$J(0, t)$, b)~$J(\ell, t)$, c)~$\log(|J(0, t)|)$ and d)~$\log(J(\ell, t))$ versus time(s). From a) and c) one can find that in a) there is a big positive peak and then five small negative ones; the first represents the wave packet as it enters the field region, the later five represent the reflected packets. From b) and d) it's concluded that there are five positive peaks in b).

Figure 6: $Q_0(t)$ (solid line) and $Q_{\ell}(t)$ (dashed line) versus time(s). The horizontal dotted line shows the  transmission probability $|T|^2$.

Figure 7: $P_0(t)$ (solid line) and $P_{\ell}(t)$ (dashed line) versus time(s).\pagebreak

%
%
\end{document}